\documentclass{emulateapj}
\usepackage{amsmath}

\shorttitle{Massive elliptical galaxies : From cores to haloes}
\shortauthors{Lintott, Ferreras \& Lahav}

\begin{document}

\title{Massive elliptical galaxies : From cores to haloes}
\author{C.J. Lintott\altaffilmark{1}, I. Ferreras\altaffilmark{1,2} \& O. Lahav\altaffilmark{1}}
\altaffiltext{1}{Department of Physics and Astronomy, University College London, Gower Street, London, WC1E 6BT, UK}
\altaffiltext{2}{Department of Physics, King's College London, Strand, London WC2R 2LS, UK}
\email{cjl@star.ucl.ac.uk}

\begin{abstract}

In the context of recent observational results that show massive ellipticals were in place at high redshifts, we reassess the status of monolithic collapse in a $\Lambda$CDM universe. Using a sample of over 2000 galaxies from the Sloan Digital Sky Survey, by comparing the dynamical mass and stellar mass (estimated from colours) we find that ellipticals have `cores' which are baryon-dominated within their half-light radius. These galaxies correspond to 3-sigma peaks in the spherical collapse model if the total mass in the halo is assumed to be 20 times the dynamical mass within the half-light radius. This value yields stellar mass to total mass ratios of 8\%, compared to a cosmological baryon fraction of 18\% derived from WMAP3 alone. We further develop a method for
reconstructing the concentration halo parameter $c$ of the
progenitors of these galaxies by utilizing adiabatic contraction. Although
the analysis is done within the framework of
monolithic collapse, the resulting distribution of $c$ is log-normal with a peak value of $c\sim3-10$ and a distribution width similar to the results of N-body simulations. We also derive scaling relations between stellar and dynamical mass and the velocity dispersion, and find that these are sufficient to recover the tilt of the fundamental plane.
\end{abstract}


\section{Introduction}
                                                                                    
The large new galaxy surveys such as 2dF \citep{2dFmain} and the Sloan Digital Sky Survey (SDSS) \citep{York} have transformed the
way we can study galaxy properties statistically.  Here we focus on a
sample of elliptical galaxies derived from
those selected from the SDSS by \citet{Bernardi1} (hereafter B03).  Our motivation is to revisit fundamental issues of galaxy
formation from the perspective afforded by a modern data set.
                                                                                    
A theory of galaxy formation in a universe dominated by Cold Dark
Matter was first set out in a seminal paper by Blumenthal, Faber,
Primack \& Rees (1986a) (hereinafter BFPR) which considers the
monolithic collapse of isolated dark matter halos. According to this
picture, the redshift at which a halo collapses is determined only by
its mass, for a given cosmology and choice of amplitude of fluctuation. The baryons follow the dark matter
distribution until the radius of the collapsing halo reaches the
virial radius and is halted.  Studies of the evolution of the
fluctuations in density, however, suggest that galaxy formation is
dominated by merging of small dark matter halos \citep{WhiteRees},
and this picture is supported by simulations \citep{Springel}.
There have been many attempts to discriminate between these
two pictures of the formation of galaxies, referred to respectively as monolithic or spherical collapse and
hierarchical merging. Simulation results notwithstanding, the small scatter of the observed colour magnitude
relation and its evolution with redshift provides evidence that massive
ellipticals were already in place at a redshift of $z\approx 1-2$ with
little subsequent merging (e.g. \citet{sed98}), whereas less massive ellipticals present features characteristic
of recent star formation (e.g. \citet{FS00}).  This
`inverted hierarchy' or `downsizing' effect \citep{Cow96}
illustrates the complexity of galaxy formation compared with a simple
picture of the assembly of dark matter halos.  In this paper we use the predictions of the spherical
collapse scenario as a benchmark, which can then be challenged with
further comparisons with observations and detailed simulations.  

A new aspect of our analysis is the estimation of the stellar mass of
each of the B03 ellipticals, allowing us to distinguish between baryonic and total mass.  We also update the
calculations of BFPR for `concordance cosmology'
(matter density $\Omega_m=0.3$,dark energy density $\Omega_\Lambda=0.7$,Hubble constant $h=0.7$ and amplitude of fluctuations $\sigma_8=0.9$) \footnote{We note that the recent WMAP release prefers $\Omega_{M}\approx 0.24$ and $\sigma_8 \approx 0.74$. However, $\sigma_8=0.9$ is still preferred by weak lensing.}
and place the observed ellipticals on a revised cooling diagram. Despite the attention BFPR has received, this relatively simple generalization to $\Lambda$CDM has not previously appeared in the literature. Our analysis of the B03 galaxies also appears to be the first attempt to incorporate data for individual galaxies (as opposed to schematic data) in this parameter space. 
                                                                               From the first three years of WMAP observations of the cosmic microwave background alone (Spergel et al. 2006), $\Omega_m=0.24^{+0.03}_{-0.04}$ and $\Omega_b=0.042^{+0.003}_{-0.005}$.
For simplicity, we use a `cosmological' baryon fraction of 1/6 unless otherwise stated.   

On galactic
scales, \citet{Klypin} model the Milky Way within the
virial radius ($\approx 250$ kpc) and find that a substantial `feedback' mechanism which removes baryons from the galaxy
itself must have operated. Do similar processes operate in massive ellipticals? Are
their cores dominated by baryonic matter?  \citet{Romanowsky} and \citet{Dekeletal} have reached conflicting
conclusions to this important question, based on the analysis of planetary nebul\ae~in elliptical galaxies out to five times the effective radius $R_e$. By using the central velocity dispersion and stellar mass out to $R_e$ we seek to answer the same question by focussing on the central region only. We find that the cores of the galaxies are baryon dominated, at least to a distance of $R_e$ from the centre.

By making the straightforward assumption that the total mass of the galaxy is proportional to the mass of the galaxy `cores' we have studied, we reproduce the BFPR results for a modern `concordance' cosmology, and investigate the regime in which their benchmark model of spherical collapse is consistent with the data. We also recover the profile of the undisturbed dark matter halo from present-day observables via the procedure of adiabatic contraction, which enables us to compare our derived profile concentration with both simulations and observations. Defining $\alpha=M_{\rm tot}$/$M_{\rm dyn}\left(<R_e\right)$ - the total mass of the collapsing halo divided by the dynamical mass within $R_e$ -- we find a strong constraint that $\alpha \ge 10$. $\alpha \approx 20$ provides a good fit to the data.

The outline of the paper is as follows: Section 2 presents the
B03 sample and a volume-limited subset of 2040
galaxies selected from it.  Section 3 describes the derivation of
stellar masses of these ellipticals from the SDSS colours.  Section 4
derives dynamical masses and Section 5 covers the relations
between luminosity, velocity dispersion, stellar mass and dynamical
mass, extending the classic \citet{FJ76} relation.  In
Section 6 we comment on the acceleration of stars in the ellipticals
and the connection to MOND. Section 7 extends the benchmark spherical
collapse model for the concordance $\Lambda$-CDM cosmology and use it to
contrast the observed ellipticals with the level of rms fluctuations on the
cooling diagram.  Finally, in Section 8 we present a new reconstruction method
which recovers the initial halo of dark matter from the present-day
observables using the adiabatic contraction.  We summarize the results
and outline future work in Section 9.

\section{The Sample}
We use galaxies selected from the sample of \citet{Bernardi1} extracted from the SDSS. \footnote{Funding for the creation and distribution of the SDSS Archive has been provided by the Alfred P. Sloan Foundation, the Participating Institutions, the National Aeronautics and Space Administration, the National Science Foundation, the U.S. Department of Energy, the Japanese Monbukagakusho, and the Max Planck Society. The SDSS Web site is http://www.sdss.org/. The SDSS is managed by the Astrophysical Research Consortium (ARC) for the Participating Institutions. The Participating Institutions are The University of Chicago, Fermilab, the Institute for Advanced Study, the Japan Participation Group, The Johns Hopkins University, the Korean Scientist Group, Los Alamos National Laboratory, the Max-Planck-Institute for Astronomy (MPIA), the Max-Planck-Institute for Astrophysics (MPA), New Mexico State University, University of Pittsburgh, University of Portsmouth, Princeton University, the United States Naval Observatory, and the University of Washington.} The sample comprises almost 9,000 early-type galaxies, with spectroscopic redshifts
in the range $0.01\le z\le 0.3$ (in the heliocentric reference frame). In B03, the authors describe the selection
criteria for their sample and the catalogue itself; as the photometry
in each of the bands is used in our subsequent analysis, it is
important to note that these criteria do not include selection by
colour. Their sample is magnitude-limited with all galaxies at a redshift of
$z>0.3$ excluded.  For the work presented in this paper we have used
the data as presented in B03, with the exception of photometry where
we have used the recent Sloan Data Release 4, \citep{SDSSDR4} which resolves many of the problems with photometry
identified in earlier releases.

\begin{figure}
\epsscale{1.0}
\plotone{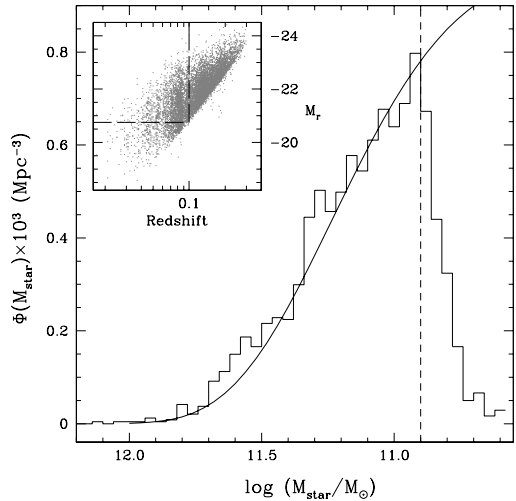}
\caption{
Stellar mass function (MF) of our volume-limited sample of early-type galaxies extracted from Bernardi et al. (2003). As shown in the insert (in which we plot the full sample of $\sim 8700$ galaxies), we initially make cuts in redshift and absolute magnitude, plotting the MF for the remaining galaxies as the jagged line. In order to check that these cuts produce a useful sample, we compare the result with the luminosity function derived for all SDSS early-type systems (type $0<T<1$) obtained independently by Nakamura et al (2003). For the purposes of this comparision, we assume a characteristic $M_{\rm star}/L_r=4$ (see figure~\ref{fig:MasstoLight}). Our sample is complete for systems with stellar masses greater than $7.94\times 10^{10} M_\odot$ (dashed line), and so we make a final cut at this mass to leave our final sample of 2040 galaxies.}

\label{fig:massfunc}
\end{figure}

Unlike the work presented in B03 and its companion papers, we have further restricted the sample to provide a volume-limited
sample.  The mass function of the subset of galaxies selected by B03,
with $z<0.1$ and absolute magnitude in the r-band $M_r<-20.75$, is
shown in figure \ref{fig:massfunc}.


 As the figure shows, restricting our sample
to massive elipticals, with $M_{\rm star} \left(<R_e\right)>7.94\mathrm{x}10^{10}M_{\odot}$
produces a volume limited sample of 2040 galaxies with $z<0.1$ which appears free
from selection effects. Unless otherwise stated, this is the sample we
use for subsequent results. The definition of stellar mass, as shown in the figure, is given in the next section.

\section{Stellar Mass}\label{stellarmass}

For massive elliptical galaxies with little gas content, the baryonic mass is approximately equal to the stellar mass at least within the effective radius, $R_e$. 
We compute the stellar mass content within the effective radius by comparing the available SDSS
photometry with a wide range of star formation histories (SFHs). We
assume an exponentially decaying star formation rate, so that each SFH
is uniquely defined by three parameters: stellar metallicity (constant
with time), star formation timescale and formation redshift (allowed
in the range $10 > z_F >2$).  These models can represent either a
short, sharp burst of star formation or a more extended episode. For
every choice of SFH we calculate the resulting synthetic stellar
populations of \citet{BruzualCharlot}, generating a composite
spectrum. A \citet{Chabrier} initial mass function is used throughout. We use four SDSS passbands to compare the flux in each of
three colours ($g-r$,$r-i$,$i-z$) with the observations of each
source. The best fit is used to determine the stellar mass-to-light
ratio which allows us to estimate the stellar mass content of each
source from the absolute $r$-band magnitudes given in data release~4.

It seems initially that the use of the full spectral energy
distribution (SED) would produce a stellar mass content which might be
more accurate. However, we must emphasize that our results indicate
that the stellar mass to light ratio in the $r$ band does not change
much with respect to the star formation histories allowed in
early-type galaxies.  All
photo-spectroscopic observables of local early-type galaxies
correspond to old, passively evolving stellar populations \citep{B03b}, which results in a very weak dependence of the M/L for the
allowed ages and metallicities  (see the right hand panel of figure \ref{fig:MasstoLight}). The largest systematic error in the
stellar mass is due to the choice of the initial mass function (IMF),
which amounts to a factor $\sim 0.1-0.2$~dex in $\log
\left(\rm{M/L}\right)$ for a range of standard IMFs (see e.g. \citet{BruzualCharlot}).

\section{Dynamical Mass}\label{sec:mdyn}

We can use the projected size and central velocity dispersion along with 
a simple model of the dynamics of early-type galaxies to estimate the
total matter content. Assuming that the velocity dispersion and
the anisotropy of the velocity distribution do not vary with radius, 
one can solve Jeans' equation (see e.g. \citet{BT88}).

Padmanabhan et al. (2004) perform a careful analysis of the profiles of massive galaxies and find that

\begin{equation}\label{eq:mdyn}
M_{dyn}\left(<R_e\right)=A\frac{\sigma^2R_e}{G},
\end{equation}

\noindent where $R_e$ is the effective (or projected half-light) radius,
$\sigma$ is the velocity dispersion as measured at the centre and A is
a constant which depends weakly on the anisotropy of the velocity field. We adopt the Padmanabhan
value of $A\approx2.72$ throughout the present work, noting that they assume a systematic error of 30\%. The dominant source of scatter is, as Padmanabhan et al. note, measurement errors.

\begin{figure}
\plotone{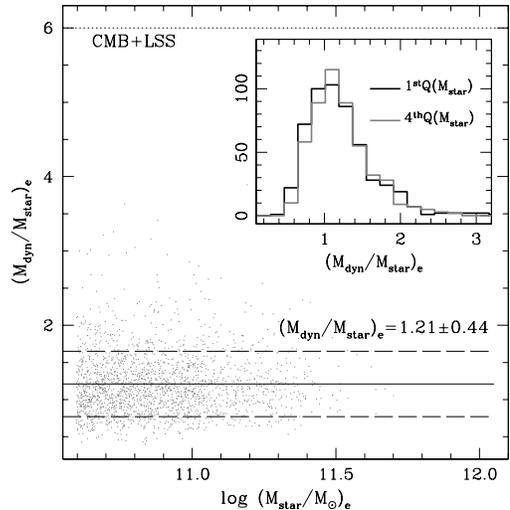}
\caption{Ratio of dynamical to stellar mass measured within R$_e$. The solid
and dashed lines give the mean and RMS of our $z<0.1$ sample. The
dotted line labelled CMB+LSS gives the prediction of the cosmological
ratio. The inset shows a histogram of the first and fourth quartiles
in stellar mass (i.e. $\log (M_{star}/M_\odot <10.72$ and $>11.02$,
respectively).}\label{fig:mdmldata}
\end{figure}

Figure \ref{fig:mdmldata} presents a comparision of the dynamical mass
and stellar mass. It is clear that dark matter is underabundant
(compared to the cosmological abundance) in the central regions of
these systems. In other words, the core region is baryon dominated, in
agreement with observations (e.g. Mamon \& Lokas 2005). The slope is in agreement with
independent estimates of total mass using strong gravitational lensing
(Ferreras et al. 2005) and suggests that dark matter is
more important in massive galaxies.  It also is comparable to the results of Loewenstein and White (1999), although they use an unrealistic model for anisotropy in their sources. 


Romanowsky et al. (2003) measure $\sigma$ through observations of bright planetary nebul\ae~ in three local ellipticals and derive a mass to light ratio within $5R_e$ compatible with those of purely stellar populations. Dekel et al. (2005) explain this result by suggesting that `the stellar orbits in the outer regions \ldots are very elongated'. Our results, derived for the innermost $R_e$ suggest that results similar to those obtained by Romanowsky et al. would be expected even in large samples of ellipticals. In addition, realistic assumptions about the velocity strucutre of the cores of these systems do not provide an explanation of the deviation observed from the cosmological baryon fraction.

\section{Scaling relations}

\begin{figure}
\plotone{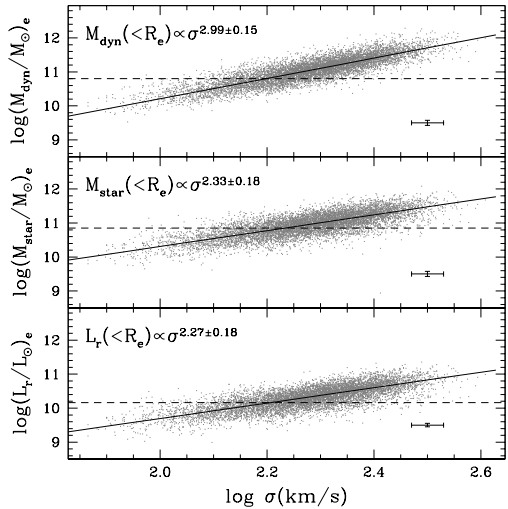}
\caption{Correlations between velocity dispersion and luminosity ({\sl bottom}), 
stellar mass ({\sl middle}) and dynamical mass ({\sl top}),
all measured within $R_e$. Characteristic error bars are shown in each
panel. The solid lines are fits to the data, with all points equally weighted. The truncation of our volume-limited sample is shown as a
dashed line. Given that the
truncation and the fit are nearly aligned, we have used the complete
sample for the fits shown in this figure.}\label{fig:msigdata}
\end{figure}

Figure \ref{fig:msigdata} shows the scaling relations of the
($r$-band) luminosity, stellar and dynamical mass as a function of
velocity dispersion. The bottom panel shows the well-known
Faber-Jackson relation \citep{FJ76}.


Notice that the truncation imposed by the choice of a volume-limited
sample results in a constraint nearly aligned with the correlation (dashed
line). Hence, a strong bias is expected if we use this reduced sample for
the fits (see e.g. appendix B in Lynden-Bell et al. 1988). Therefore we
use the complete sample for the fits in this figure.
                                                                                
In order to reduce the effect of outliers in the calculation of the best
fitting parameters, in common with all fits in this paper we use a robust
M-estimator based on the minimisation of the mean absolute deviation
(Press et al. 1992), as standard least squares methods were found to be
very sensitive to outliers in this sample.  Given the uncertainties in determining the errors in velocity dispersion and radius, we have chosen to weight all points equally rather than take into account their quoted errors. A bootstrap method is used to
determine the uncertainty in the slope. The same fitting technique was
applied on 1000 bootstrapped samples of 500 galaxies each. The predicted
slopes from each run were used to determine the quoted uncertainties.


The Faber-Jackson relation is a direct consequence of the dynamics of
the systems being studied. Early-type galaxies are hot dynamical
systems whose support comes mostly from the velocity dispersion of the
constituent stars. Hence, a correlation is expected between the
central velocity dispersion and the mass of the system.  The different
slope between stellar and dynamical mass estimates is mostly related
to either stellar population effects, to structural differences, or to
a significant correlation between the dark matter fractional
contribution and total galaxy mass.  This effect has been commonly
invoked to explain the tilt of the observed so-called `fundamental
plane' with respect to the prediction from the virial theorem (see
e.g. \citet{Guz93}; \citet{FS00b}; \citet{Truj04}).

\begin{figure}
\plotone{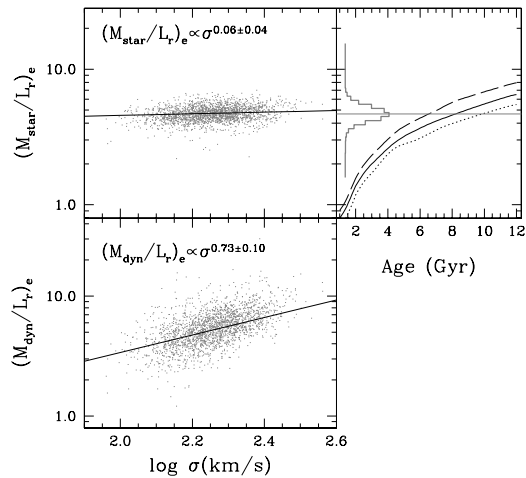}
\caption{Stellar ({\sl top}) and dynamical mass-to-light ratios as a
function of velocity dispersion for our volume-limited sample, after cuts have been applied. The straight lines are fits to the
data, and the corresponding power law indices are labelled in each
panel. The panel on the right shows the predictions for simple stellar
populations from the models of Bruzual \& Charlot (2003) as a function of 
age. The dashed, solid and dotted lines
correspond to metallicities 2, 1 and 1/2 times solar. The histogram 
refers to the data (left panel), with the average given as a horizontal
line.}\label{fig:MasstoLight}
\end{figure}

Figure~\ref{fig:MasstoLight} shows the correlation between velocity
dispersion and M/L (stellar or total).  Local early-type galaxies are
dominated by stars which are older than about $7-8$~Gyr, which implies
that the contribution of stellar populations to the tilt of the
fundamental plane cannot be too large, as previously suggested
\citep{Dr87}.


There is a remarkable difference between stellar and dynamical M/L in
figure~\ref{fig:MasstoLight}. The fits were done using a bootstrap
minimisation of the mean absolute deviation on the volume-limited sample.
The slope of the dynamical M/L is sufficient to
explain the tilt of the fundamental plane. Notice that
equation~\ref{eq:mdyn} assumes homology for all galaxies. Hence, we
recover the tilt without invoking structural
non-homology. Furthermore, recent estimates using strong gravitational
lensing \citep{FSW} suggest dark matter is the main cause of the tilt,
a view supported by these results.

\section{Acceleration}

Our analysis so far has been done within the framework of Newtonian
dynamics. We showed that the cores (i.e. within radius $R_e$)
of massive early-type ellipticals are dominated by
by baryons, rather than by dark matter.
The alternative model of Modified Newtonian Dynamics (MOND) of
\citet{Milgrom} suggests a modification to the acceleration if it is below
$a_0 = 10^{-8} \mathrm{cm s^{-2}}$.
This modification can then explain the rotation curves of spiral galaxies
without the need for dark matter.
It is intriguing indeed to see a correlation between the Newtonian
$M/L$ and the centrifugal acceleration $(a=v^2/r)$
at the last measured radial point $r$ in spiral galaxies \citep{SandersMcGaugh}. Their points are reproduced in our Figure 5 (as circles).
It is clear from the start that such low acceleration
is not expected at the cores of ellipticals which have much smaller radii
than the outer parts of spiral galaxies.
However, it is interesting to look for any such correlation in the sample
of ellipticals. Figure 5 shows (in dots) the dynamical $M/L_K$ vs. the
acceleration, defined as $a=v^2/R_e$ = $3 \sigma^2/R_e$.
We see no obvious correlation and no continuity between the spirals and
the ellipticals.
As expected the ellipticals have accelerations which are
$\sim 10 $~times larger than that required by MOND to show departure form
the Newtonian dynamics. If indeed ellipticals have large halos it would be
an important test to measure the acceleration at these large radii.


\begin{figure}
\plotone{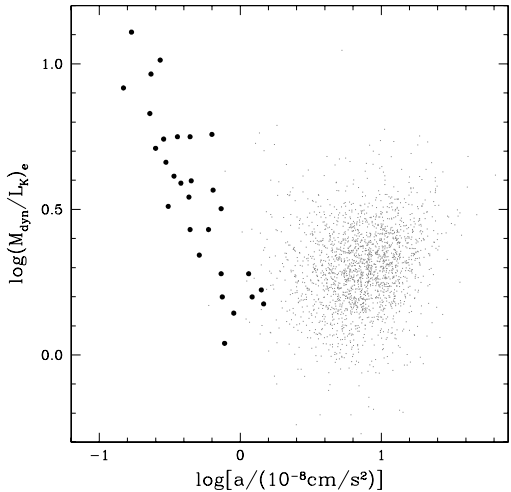}
\caption{Comparison of the acceleration of early-type and late-type galaxies.
We plot the dynamical M/L$_K$ as a function of acceleration (defined as
$a=v^2/R_e=3\sigma^2/R_e$. In order to generate $K$ band M/L ratios, we use 
$r-K=2.4$, typical of early-type systems. The late-type data (circles)
come from Sanders \& McGaugh(2002). Notice that early-type galaxies fall
outside of the low-acceleration regime required by MOND to give a 
significant departure from Newtonian dynamics.}\label{fig:acc}
\end{figure}

\section{Spherical Collapse in a $\Lambda$CDM Universe}

We have already mentioned the renewed interest in spherical collapse models as part of the complexity of galaxy formation. How do the predictions of this simple model compare to the modern data we have available, and how do the predictions change in a cosmological model incorporating a cosmological constant? There have been many attempts to study spherical collapse in $\Lambda$CDM (e.g. Lahav et al. 1991; Wang \& Steinhardt 1998) but we will follow the analysis of BFPR more closely, using the velocity dispersion-mass and baryon density-temperature plots to compare the predictions of the model with data. 

\begin{figure}
\plotone{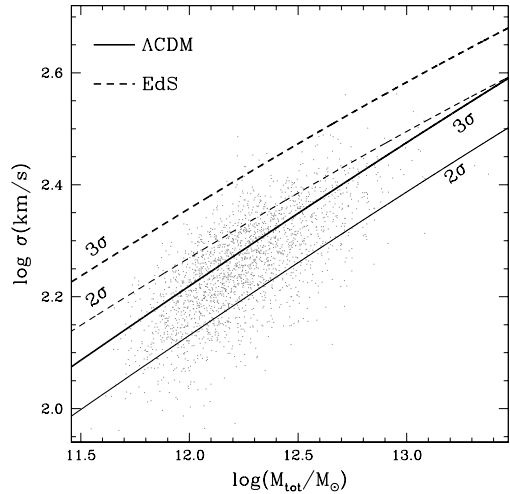}
\caption{A comparision of the spherical collapse model with our
sample of massive early-type galaxies for both EdS (dashed lines) and $\Lambda$CDM cosmologies. (EdS parameters : $\Omega_m=1.0$,$\Omega_\Lambda=0.0$,$h=0.7$,$\sigma_8=0.9$, $\Lambda$CDM parameters : $\Omega_m=0.3$,$\Omega_\Lambda=0.7$,$h=0.7$, $\sigma_8=0.9$). In obtaining $M_{\rm tot}$ we have assumed $\alpha=20$ (see text for details). The $M_{\rm tot}$ obtained are consistent with a formation process from 2- or 3-sigma perturbations in a $\Lambda$CDM cosmology.}\label{fig:newmv}
\end{figure}

In order to check whether a simple spherical collapse model is compatible with the properties of massive galaxies, we show in figure \ref{fig:newmv} the correlation between total mass (i.e. including the halo) and the central velocity dispersion. The lines represent the
theoretical mass and velocity dispersion of dark matter haloes produced by
monolithic collapse. The collapse of a spherical, top-hat,
perturbation is described by:

\begin{equation}\label{eqn:collapse}
\frac{\mathrm{d}^2r}{\mathrm{d}t^2}=H_0^2\Omega_\Lambda r - \frac{GM\left(<r\right)}{r^2},
\end{equation}

\noindent where r is the radius in a fixed coordinate frame, $M\left(<r\right)$ is the mass
enclosed within r and the initial velocity field is equivalent to the
Hubble flow only. If the mass described by such an equation is bound,
it will expand initially before reaching a maximum radius and
collapsing. Linear theory predicts that the over-density at which the top
hat collapses to a point is $\delta_{\rm linear}=1.686$
\citep{Peebles}. This value is appropriate for an Einstein-de Sitter
universe, but the value is only slightly sensitive to the choice of
cosmology. Thus, for a top hat to be on the point of collapsing at a
redshift $z$, its density extrapolated to the present day must satisfy:

\begin{equation}\label{eqn:deltac}
\delta_{c}\left(z\right)=\frac{1.686}{D\left(z\right)}
\end{equation}

\noindent where $\rm{D\left(z\right)}$ is the growth factor from linear theory. 

Collapse to a single point in this manner is prevented by violent
relaxation (`phase mixing') which produces a state of virial
equilibrium. \cite{BryanNorman} use the virial theorem to obtain the
final over-density (relative to the critical density at the collapse
redshift) for universes in which $\Omega_m + \Omega_\Lambda =1$:

\begin{equation}\label{eqn:bndeltac}
\Delta_c=18\pi^2 + 82d -39d^2
\end{equation}

\noindent where $d=\Omega_m^z - 1$ and

\begin{equation}
\Omega_m^z =
\frac{\Omega_m\left(1+z\right)^3}{\Omega_m\left(1+z\right)^3+\Omega_\Lambda+\Omega_k\left(1+z\right)^2}.
\end{equation}

\noindent A halo will therefore collapse at a redshift, z, determined only by its total mass and by cosmology.
Via the virial theorem its radius will be given by \citep{B&L}:

\begin{eqnarray}\label{eq:rvir}
r_{vir}=0.784\left(\frac{M_{tot}}{10^8h^{-1}M_{\odot}}\right)^{\frac{1}{3}}\left[\frac{\Omega_m}{\Omega_m^z}\frac{\Delta_c}{18\pi^2}\right]^{-\frac{1}{3}}& &\nonumber \\
{}* \left(\frac{1+z}{10}\right)^{-1}h^{-1} \mathrm{kpc}. & &
\end{eqnarray}

\noindent A halo of given mass will collapse later in a universe with a non-zero
cosmological constant, leading to the differences between the two sets
of calculations presented here. The Einstein-de Sitter Universe
calculations were first presented by BFPR, and here we have updated
their diagram to reflect a modern cosmological model together with data for individual galaxies which, to the best of our knowledge, has not been done before. 

In order to compare with data, it is necessary to convert the
$M_{\rm dyn}$ defined by equation \ref{eq:mdyn} to a total mass. A naive
solution would be to assume that $M_{\rm tot}\approx 2M_{\rm dyn}\left(<R_e\right)$ but this
ignores factors including variation in the mass to light ratio and the
profile of the velocity dispersion. Instead, we use the ansatz that
the total mass of a galaxy will be proportional to the dynamical mass
within $R_e$:

\begin{equation}
M_{\rm tot}=\alpha M_{\rm dyn}\left(<R_e\right)
\end{equation}

There will exist a scatter in $\alpha$ in the population of galaxies, and we may expect $\alpha$ to be a function of the total mass of the galaxy. However, if we approximate by considering a single value of $\alpha$, we can then consider for what values of $\alpha$ the spherical
collapse model discussed above is appropriate. Consider the example of
$\alpha=2$, the naive value discussed above. The slope of the data is
encouraging, being in good agreement with the theoretical
prediction. However, for a given mass, sigma is so large that the
galaxies correspond to 4-sigma fluctuations in an EdS model and a
7-sigma fluctuation in $\Lambda$CDM. These are extremely unlikely fluctuations, not compatible with the observed number density of massive galaxies.

In order to get a rough estimate for the level of fluctuations that
correspond to massive galaxies, we use the morphologically segregated
luminosity functions (LFs) from SDSS (Nakamura et
al. 2003). Integrating the LF of E/S0s from L$_\star$ to some bright
upper limit, say 20L$_\star$ (the choice does not matter given the
exponential cutoff of the LF), we get a comoving number density of
ellipticals similar to those in our sample (see
figure~\ref{fig:massfunc}). The numbers from the SDSS LF give
$n($E/S0$)=0.11\times 10^{-2}h^3$Mpc$^{-3}$. A similar integral for
the LF of {\sl all} galaxies in the sample in the range
$0.005<$L/L$_\star < 20$ gives the comoving number densities of
galaxies, $n($All$)=9.02\times 10^{-2}h^3$Mpc$^{-3}$. The ratio of the two 
number densities ($n($E/S0$)/n($All$)=0.012$) corresponds to a Gaussian
fluctuation around 2.5-sigma. Hence, we can assume that a spherical
collapse model from 2- to 3-sigma level fluctuations is compatible with the
observed number density of early-type galaxies today. A choice of $\alpha=20$ produces a good
correlation between the data and the predicted M-$\sigma$ relation for
3-sigma fluctuations given $\Lambda$CDM.

It is interesting to note that this choice of $\alpha$ produces a
value for the stellar mass to total mass ratio which is approximately half
the cosmological baryon fraction; i.e. half the baryons in the halo are found in the form of stars. Models with large $\alpha$ are
also consistent with the large mass to light ratios found by studies
of massive ellipticals in x-rays \citep{Griffiths}.

\section{The Evolution of the Dark Matter} \label{sec:adiab}

In order to further investigate the distributions of total and stellar
mass within the spherical collapse framework, we need to take into
account the evolution of the dark matter as the baryons dissipate
energy and sink towards the center of the halo. We use the procedure
of adiabatic contraction of the dark matter halo, following \citet{BFFP}.  We assume that the dark matter has an initial density
profile of the form given by \citet{NFW} (hereafter
NFW).  The concentration is defined as $c=r_{\rm vir}/r_s$ where
$r_{\rm vir}$ is the virial radius -- usually considered to be the
edge of the halo -- and $r_s$ is the scale radius. This method is
strictly valid only for halo particles moving on circular
orbits. \citet{Gnedin} show that the predicted density profile is more
dense than that found in N-body simulations. However, the deviation is small \citep{SellwoodMcGaugh}
and we use the method here as a first approximation.

The initial mass profile and the final dark matter and baryonic
profiles are related by two equations:
\begin{equation}
r\left[M_g\left(<r\right) + M_h\left(<r\right)\right]=r_iM_i\left(<r_i\right),
\end{equation}
\begin{equation}
M_h\left(<r\right)=\left(1-f_{cool}\right)M_i\left(<r_i\right).
\end{equation}

\noindent $M_i\left(<r\right)$ is the initial mass enclosed within the radius,r. $M_g\left(<r\right)$ and $M_h\left(<r\right)$
are the final baryonic and dark matter profiles
respectively. $f_{cool}$ is the fraction of the system's mass which is
contained in baryons which cool to form the galaxy, and is related to
$\alpha$ via the relation:

\begin{equation}\label{eq:fcool}
f_{cool}=\frac{2M_{star}\left(<R_e\right)}{\alpha M_{dyn}\left(<R_e\right)}
\end{equation}

\noindent where the stellar mass within $R_e$ is assumed to be half the cooled baryonic mass
(\citet{Padh}). For $\alpha=20$ and the ratio of
$M_{star}\left(<R_e\right)$ to $M_{dyn}\left(<R_e\right)$ found in
section \ref{sec:mdyn} a value of $f_{cool}\approx 0.08$. This corresponds to half the cosmological baryon fraction, which suggests that half the halo's baryons are in stars. We can also use this sort of argument to place a limit on $\alpha$ by requiring $f_{\rm cool}<$1/6 (where 1/6 is the cosmological baryon density). This corresponds to $\alpha>10$.

The baryons are assumed to have a density profile of the form given by
\citet{Hernquist} (which closely approximates the de Vaucouleurs
law, see \S4). In this case the adiabatic contraction problem can be solved
analytically \citep{Keeton}. Each initial radius $r_i$ maps to a final radius $r$ given
by:
\begin{equation}
f_{cool}r^3+\left(r+a\right)^2\left[\left(1-f_{cool}\right)r-r_i\right]m_i\left(<r_i\right)=0,
\end{equation}
\noindent
where $a=0.551R_e$ is the scale radius of the Hernquist profile, and
$m_i(<r_i)=M_i(<r_i)/M_{\rm tot}$. 

A halo of mass $M_{tot}$ in the
monolithic picture considered here will collapse at a redshift determined only by the cosmology and the amplitude of perturbations. The virial radius is then given by equation
\ref{eq:rvir}. After choosing a value for $\alpha$, $f_{\rm cool}$ is given by equation \ref{eq:fcool} using the value of $M_{\rm star}\left(<R_e\right)$/$M_{\rm dyn}\left(<R_e\right)$ for each galaxy.  For any concentration, c, we
then obtain the total mass inside the initial radius ($r_i$) that will evolve
into a final radius $r=R_e$. The adiabatic contraction prescription implies this mass
is equal to the final mass observed within $R_e$. Hence, we determine via a 
maximum likelihood estimator the concentration that satisfies $M_{\rm dyn}\left(<R_e\right)=M_i\left(<r_i\right)$.

\subsection{Results}

\begin{figure}
\plotone{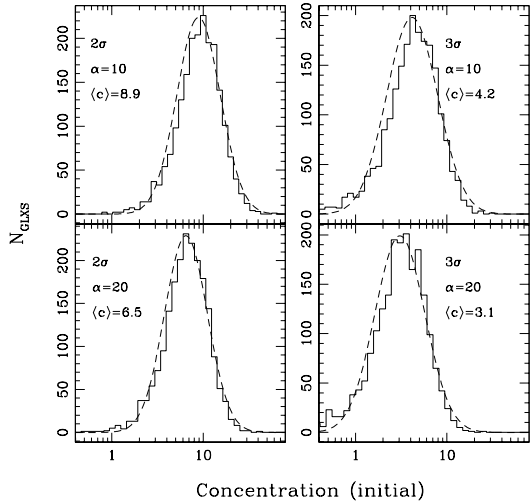}
\caption{Concentration of the (NFW) dark matter halos {\sl before} adiabatic
contraction.  The benchmark spherical collapse model, assuming LCDM
and either 2- or 3-sigma perturbations is used. $\alpha=M_{\rm tot}$/$M_{\rm dyn}\left(<R_e\right)$ is either 20 or 10 as indicated. In all cases an
approximately log-normal distribution of concentration is recovered. A
higher order perturbation (3- instead or 2-sigma) or larger
$\alpha$ reduces the mean value of the concentration.}\label{fig:c}
\end{figure}

Figure~\ref{fig:c} shows the histogram of the concentration of the
{\sl initial} DM halos for our sample, for 2 and 3-sigma
perturbations and $\alpha$ of 10 or 20. The figure overlays a Gaussian
distribution corresponding to this result. The fit is remarkable:
our simplistic model of spherical collapse and
adiabatic contraction combined with a volume limited sample, results
in a log-normal distribution of concentrations.


This type of distribution is also obtained in more
detailed N-body simulations \citep{Bll01}. Furthermore, the
dispersion of our distribution is $\sigma (\log c)\sim 0.2-0.25$, in
agreement with the simulations, which give $\sim 0.18$;  notice that these simulations give values of concentration higher than
those we find even for 2-sigma fluctuations. 

Hierarchical models give concentrations around 4 only for objects which have recently undergone major mergers or for large (galaxy cluster mass) systems. The mean concentration for galaxies with mass $M_{\rm vir} \ge 1.5{\rm x}10^{12}h^{-1}M_{\odot}$ is typically 13.1 (Wechsler et al. 2002).

As figure
\ref{fig:c} shows, our model does reproduce large values of c for
small values of $\alpha$ and 2-$\sigma$ fluctuations. However, a value
$\alpha < 10$ would shift the data points in figure \ref{fig:newmv} to the
left, away from the predictions for 2- or 3-sigma fluctuations. We
therefore conclude that the benchmark model predicts a mean
concentration lower than that found in simulations.

These results are reminiscent of those of \citet{Collister}, who studied the concentration of galaxy clusters and groups in the 2dF galaxy redshift survey. They found concentrations from the galaxy data which were a factor of $\sim$2 lower than the predictions of dark matter simulations (e.g. \citet{Kravtsov}), albeit on larger scales than those considered here. Trujillo et al. (2006) present weak evidence for a change in concentration of massive elliptical galaxies (as derived from the size-mass relation) with redshift; further studies with larger samples should enable us to test their results with our methods.

\subsection{Galaxy formation and cooling}

\begin{figure}
\plotone{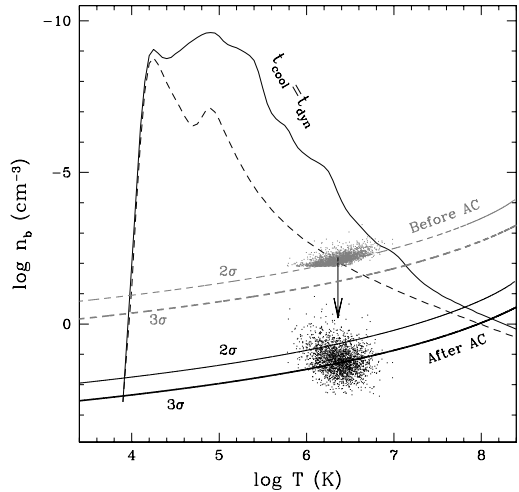}
\caption{Galaxy formation {\sl \`a la BFPR} Our sample galaxies are
shown in a density-temperature diagram. The black dots represent the observed sample,
whereas the grey dots ``extrapolate'' the sample to the time of turnaround, before
adiabatic contraction ensued. The grey and black lines are predictions for the collapse of uniform spheres corresponding to $2$ and $3\sigma$ fluctuations, and are also shown before
and after adiabatic contraction (AC). See text for details on the cooling curve; the dashed line corresponds to zero metallicity, the solid line to solar.}
\label{fig:cool}
\end{figure}

\begin{figure}
\plotone{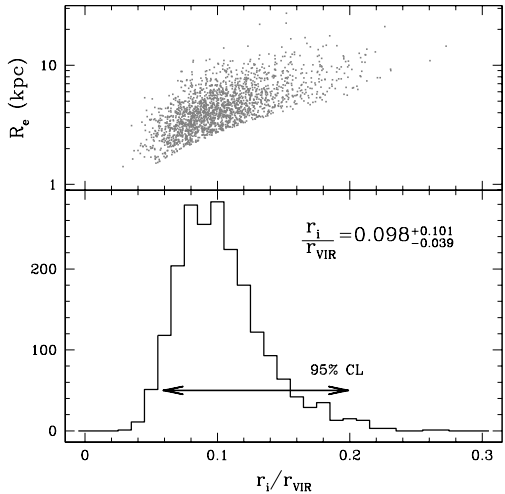}
\caption{Comparision of the virial radius $r_{\rm vir}$ and the initial radius used in the adiabatic contraction procedure. For each galaxy $r_i$ is chosen such that the mass within $r_i$ before contraction corresponds to the dynamical mass within $R_e$ after contraction. The top panel shows the correlation with the effective radius, $R_e$. In larger galaxies, more of the halo is contracted into $R_e$.}\label{fig:rf}
\end{figure}

In order to consider the baryonic component of the galaxies, it is
convenient to plot baryon number density against temperature, following
BFPR. Defining number density at present as:

\begin{equation}\label{eq:nb}
n_b=\frac{3M_{\rm star}(<R_e)}{4\pi R_e^3\mu m_H}
\end{equation}

\noindent 
and temperature as:

\begin{equation}
T=\frac{\mu m_H\sigma^2}{k_B},
\end{equation}

\noindent where $\mu m_H$ is the mean atomic weight ($\mu=0.6$ for a primordial
mix of hydrogen and helium), and $k_B$ is the Boltzmann constant. The
result is shown in figure \ref{fig:cool}. The sample is shown in two different ways: the black dots are the actual baryon densities (i.e. after
adiabatic contraction) using equation \ref{eq:nb}, computed within the
half-light radius.
The grey dots represent baryon densities before adiabatic contraction, and are
computed as the density corresponding to a baryon mass $\langle f_{cool}\rangle M_{\rm tot}$
within a radius $R_{\rm vir}$. The grey and black lines
are predictions for the spherical collapse model ($\Lambda$CDM), and are shown for 2 and
3-sigma fluctuations before and after adiabatic contraction. These models require
a translation from total mass to baryon mass, for which we use the average value
for $f_{cool}$ obtained from the sample, i.e. $\langle f_{cool}\rangle = 0.083$. For the `before AC' (grey) lines we use this mass and the virial radius definied by equation \ref{eq:rvir}. Incidentally, we notice that the early-type galaxies indicated in 
the original cooling diagram in BFPR corresponds to our 
set of points before AC. As far as we know, this is the first time individual galaxies have been plotted in this space as previous work has relied on `schematic' data. The (black) curves ``after AC'' also need to assume a ratio between the virial radius and
the initial radius ($r_i$) which is mapped into $R_e$. We use again the 
sample to find $r_i\sim 0.1 R_{vir}$ (see figure \ref{fig:rf}). The figure shows consistency with our previous 
results. 

The curves tracing $t_{dyn}=t_{cool}$ uses the dynamical timescale:

\begin{equation}
t_{\rm dyn}=\sqrt{\frac{3\pi}{32G\rho}}
\end{equation}

where $\rho$ is the total density of baryonic and dark matter and the cooling timescale is 

\begin{equation}
t_{\rm cool}=\frac{3k_BT}{n\Lambda_{\rm net}}
\end{equation}

where n is the number density of nuclei and $\Lambda_{\rm net}$ is the net cooling rate for solar (or zero) metallicity from \citet{sd93}. As expected, the galaxies after contraction lie in the region of the plot where $t_{\rm cool}<t_{\rm dyn}$. Although the approach is simplistic, assuming that galaxies are uniform spheres rather than using radial profiles, it is intriguing that 2-sigma fluctuations even for massive galaxies lie on the cooling curve for zero metallicity systems, and that adiabatic contraction promotes cooling of these systems. This result may be followed up with detailed simulations.

\section{Conclusions}

The sample of early-type galaxies selected by Bernardi et al. (2003a)
from the Sloan Digitial Sky Survey has been used to investigate the
properties of baryonic and dark matter within these systems. For the
work presented in this paper, the sample was further restricted to be
volume limited, containing only massive elliptical systems, leaving a
total of more than 2000 systems. Using photometry from Sloan Data Release
4 and the synthetic stellar populations of Bruzual \& Charlot (2003) we
calculated the mass in stars of each system.

We recover the well-known Faber-Jackson relation between luminosity
and observed velocity dispersion and also find a tight correlation
between stellar mass and velocity dispersion, which we have called the
`baryonic Faber-Jackson relation'. We find a slope of  $M_{\mathrm{star}}\left(<R_e\right) \propto \sigma^{2.33 \pm
0.18}$ (in appropriate units), less steep than previous work
\citep{Thomasetal} based on a study of an order of magnitude fewer
galaxies.

In addition to determining the stellar mass, it is possible to
determine a dynamical mass via the velocity dispersion and the radius
of the system. Comparing the two estimates of mass reveals that dark
matter is underabundant in the centre of such galaxies when compared
to the cosmological baryon fraction. However, our results are in good agreement with observers
who have claimed that the inner regions of massive elipticals are
baryon dominated.

In order to examine whether this large data set is sufficient to
discriminate between galaxy formation via monolithic collapse or the
merging of small dark matter haloes, the data were compared to a
simple model of spherical collapse. It was necessary to assume that
the total mass of the systems under study is proportional to the mass
within the baryon-rich core (within $R_e$), but given such an assumption we find that
for $\alpha \approx 20$ (where $\alpha=M_{\mathrm{tot}}$/$M_{\mathrm{dyn}}\left(<R_e\right)$),
the data were found to be consistent with simple monolithic collapse
of 3-sigma peturbations (assuming $\Lambda$CDM cosmology).

Interestingly, this choice of $\alpha$ also produces a value for the fraction of the mass which lies in the form of cooled baryons which is consistent with other work. A choice of 20 seems large, but corresponds to the stellar mass contributing half of the cosmological baryon fraction. Given the other reservoirs of baryons in these systems, this is a reasonable value.   

Having motivated our choice of $\alpha$, we used the adiabatic
contraction method first due to \citet{BFFP} to calculate the response
of the dark matter to the presence of the observed baryons. We assume
that the dark matter has an initial NFW profile, and produce the
distribution of initial concentration shown in figure \ref{fig:c}. The
mean concentration is lower than that found in dissipation-less
N-body simulations.

In summary, we have shown that the B03 sample of early-type galaxies
is a powerful resource for studying the distribution of matter within
these objects. However, despite this excellent data set we cannot rule
out spherical collapse as a mode of formation for massive
ellipticals. We find that the central regions of the
galaxies, which are baryon dominated, must represent only a small
fraction of the total mass of the system. In future work, we will
extend our analysis to a new large sample  of
early-type systems, and undertake a detailed comparision of
simulation results in order to understand the differences between
observations and semi-analytic models.

\section*{Acknowledgements}
 
We would like to thank Jacob Bekenstein, Sandra Faber and Rachel Somerville for valuable discussions. OL is supported by a PPARC Senior Fellowship, and CJL by a PPARC Studentship.

\clearpage


%

%

%

%

%

%
%


\label{lastpage}

\end{document}